\documentclass[doublecol]{epl2} 
\usepackage{amssymb}
\usepackage{amsfonts}
\usepackage{amsmath}
\usepackage{amsthm}
\usepackage{bbm}

\title{Delayed-feedback chimera states: Forced multiclusters and stochastic resonance}
\shorttitle{Delayed-feedback chimera states: Forced multiclusters and stochastic resonance} %Insert here a short version of the title if it exceeds 70 characters

\author{V.~Semenov\inst{1} \and A.~Zakharova\inst{2} \and Y.~Maistrenko\inst{2,3} \and E.~Sch{\"o}ll\inst{2}}
\shortauthor{V.~Semenov \etal}

\institute{                    
  \inst{1} Department of Physics, Saratov State University, Astrakhanskaya str. 83, 410012 Saratov, Russia\\
  \inst{2} Institut f{\"u}r Theoretische Physik, Technische Universit\"at Berlin, Hardenbergstra\ss{}e 36, 10623 Berlin, Germany\\
  \inst{3} Institute of Mathematics and Center for Medical and Biotechnical Research, NAS of Ukraine, Tereschenkivska Str. 3, 01601 Kyiv, Ukraine\\
}

\abstract{A nonlinear oscillator model with negative time-delayed feedback is studied numerically under external deterministic and stochastic forcing. It is found that in the unforced system complex partial synchronization patterns like chimera states as well as salt-and-pepper like solitary states arise on the route from regular dynamics to spatio-temporal chaos.
The control of the dynamics by external periodic forcing is demonstrated by numerical simulations. It is shown that one-cluster and multi-cluster chimeras can be achieved by adjusting the external forcing frequency to appropriate resonance conditions. If a stochastic component is superimposed to the deterministic external forcing, chimera states can be induced in a way similar to stochastic resonance,
they appear, therefore, in regimes where they do not exist without noise.}

\begin{document}

\maketitle
\section{Introduction}
Chimera states are complex spatio-temporal patterns in ensembles of identical oscillators, composed of coexisting domains of coherent (synchronized) and incoherent dynamics \cite{KUR02a,ABR04}. Chimera states were studied in detail both theoretically as reviewed in \cite{MOT10,PAN15} and experimentally \cite{HAG12,TIN12,MAR13,WIC13,SCH14a,ROS14a,KAP14,GAM14}. Only recently, the deliberate control of chimera patterns has been considered \cite{SIE14c,BIC15,OME16}. In real-world systems chimera states might play a role, e.g., in the unihemispheric sleep of birds and dolphins~\cite{RAT00}, in neuronal bump states~\cite{LAI01,SAK06a}, in epileptic seizure \cite{ROT14}, in power grids~\cite{MOT13a}, or in social systems~\cite{GON14}. The influence of noise upon chimera states is also of interest, because fluctuations are inevitably present in all real-world systems. Noise can lead to absolutely opposite effects: either destroy deterministic dynamics or increase the temporal coherence as in the case of coherence resonance \cite{HU93a,PIK97,LIN99a,LIN04} and stochastic resonance \cite{GAM98, ANI99}. While the question of robustness of chimera states with respect to random fluctuations has been considered previously \cite{LOO16}, the constructive role of noise for chimera states remains to be understood. 

Chimera states, which were initially revealed and investigated in ensembles of coupled oscillators, have also been found in single oscillators with time-delayed feedback \cite{LAR13,LAR15}. It is well-known that in the presence of time delay simple dynamical systems can exhibit complex behavior, such as delay-induced bifurcations \cite{SCH03h}, delay-induced multistability \cite{HIZ07}, stabilization of unstable periodic orbits \cite{PYR92} or stationary states \cite{HOE05}, to name only a few examples. As noted in \cite{ARE92}, there exists an analogy between the behavior of time-delayed systems and the dynamics of ensembles of coupled oscillators or spatially extended systems \cite{GIA96,MAR12a}. Certain spatio-temporal phenomena (for example coarsening \cite{GIA12}) can be tracked down in the purely temporal dynamics of time-delay system by using this approach, which considers the delay interval $[0, \tau]$ in analogy with the spatial coordinate.  Chimera states in time-delayed feedback oscillators are manifested as a sequence of regular dynamics (coherent domain) and chaotic dynamics (incoherent domain) during each time interval 
$[0, \tau]$ of the time series. 

In the present work chimera states are explored in a time-delay system which is similar to the Ikeda model with time-delayed feedback 
\cite{IKE79}. However, in contrast to the Ikeda model with positive feedback studied in \cite{LAR13,LAR15}, here we consider negative time-delayed feedback.  As we will show in the following, such negative feedback results in a different scenario leading from complete coherence to complete incoherence, and moreover, if the system is forced by applying external deterministic and stochastic signals, novel phenomena
like stochastic resonance of chimeras arise. Thereby we link two effects which have been studied independently before: stochastic resonance and chimera states. The motivation for applying negative feedback comes on the one hand from the general observation in networks that positive (excitatory) and negative (inhibitory) couplings, respectively, often lead to completely different dynamic behavior, for instance in neuroscience \cite{VRE94,LEH11,KEA12,LAD13}, and on the other hand from the possibility of experimental realization as an electronic circuit.

\section{Model}
We consider the following paradigmatic nonlinear delayed feedback oscillator:
\begin{equation}
\label{eq1}
\varepsilon \dot{x} = -y-gx-f(x(t-\tau)), \quad
\dot{y}=x-Q(y)
\end{equation}
where $x$ and $y$ are the fast and slow variable, respectively, $\varepsilon\ll 1$ is the time scale ratio, $g>0$ is a damping parameter, $Q(y)$ characterizes the oscillator nonlinearity, and the nonlinear function $f(x)$ represents the time delayed feedback with delay time $\tau$. 

Such a system describes, for instance, an electronic circuit with two nonlinear elements, where $x$ and $y$ are dimensionless voltage and current, respectively, the first Eq.~(\ref{eq1}) is the current balance, and $g^{-1}$ is the linear resistance. The function $f(x)$ is an approximation of the current-voltage characteristic of, e.g., the lambda-diode based circuit \cite{KAN75}, and the function $Q(y)$ can describe current-controlled negative resistance.
A simple realization is given by:
\begin{equation}
\label{functions}
f(x)=\frac{x}{ax^{2}+b}, \quad
Q(y)=
\begin{cases}
         -m_{1} y , & y < 0,\\
         -m_{2} y, & y\geq0.
\end{cases}
\end{equation}
and $a,b,m_1,m_2$ are positive parameters.

%%%%%%%%%%%%%%%%%%%%%%FIG4%%%%%%%%%%%%%%%%%%%%%%%%%%%%%
\begin{figure}[]
\begin{center}
\includegraphics[width=0.45\textwidth]{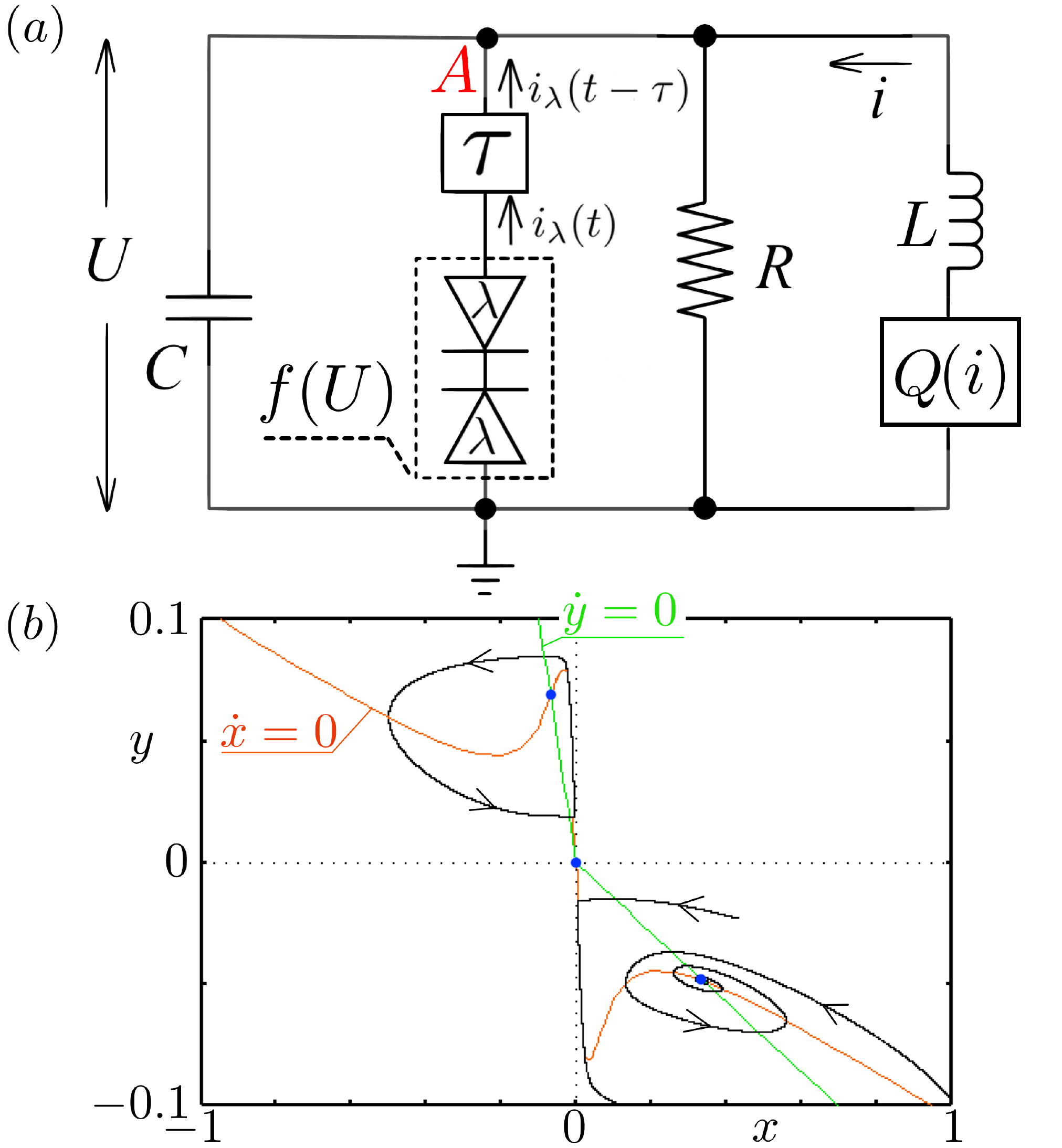}
\end{center}
\caption{(a) Scheme of electronic circuit. (b) Phase portrait in the ($x,y$) plane of system (\ref{eq1}): The $\dot{x}$ and $\dot{y}$ nullclines are shown in red (dotted) and green (dashed), respectively; stable limit cycle (black with arrows) on the left branch of the $\dot{x}$ nullcline, fixed points (blue): stable focus on the right branch, unstable focus on the left branch, saddle-point in the origin. Parameters: $\varepsilon = 0.005, a = 200,b = 0.2, g = 0.1,m_{1} = 7, m_{2}=1, \tau=0$.}
\label{fig4}
\end{figure}
%%%%%%%%%%%%%%%%%%%%%%%%%%%%%%%%%%%%%%%%%%%%%%%
The electrical circuit which is an exemplary realization of the delayed-feedback oscillator Eq.~(\ref{eq1}) is shown schematically in Fig.~\ref{fig4}a. It is a self-oscillatory circuit with parallel resistance $R$, capacitance $C$, and inductance $L$, including two nonlinear elements $f(U)$ and $Q(i)$ and time-delay $\tau$. $U$ is the voltage and $i$ is the current, and $f(U)$ models the feedback term represented by a lambda diode, whose current-voltage characteristic can be approximated by the form $i_{\lambda}(U_{\lambda})=\frac{U_{\lambda}}{aU_{\lambda}^2+b}$ with parameters $a, b>0$. The second nonlinear element $Q$ in Fig.~\ref{fig4}a is a current-controlled negative resistance which has a piecewise linear voltage-current characteristic with two slopes $m_1, m_2>0$
\begin{equation}
\label{negative}
U_{Q}(i)=
\begin{cases}
         -m_{1} i , & i < 0,\\
         -m_{2} i, & i\geq0.
\end{cases}
\end{equation}

By using Kirchhoff's laws for the node A  (see Fig. \ref{fig4}a) the differential equations (\ref{eq1}), (\ref{functions}) can be derived,
where $x$ is the voltage $U$ across the capacitor $C$, $y$ is the current $i$ through the inductor $L$, $\varepsilon=\frac{C}{L}, g=\frac{1}{R}$, and time has been rescaled by $t/L$. 

Without time delay ($\tau=0$) Eq.~(\ref{eq1}) is a bistable oscillator which exhibits two coexisting attractors (limit cycle and fixed point) in the phase space, as depicted in Fig.~\ref{fig4}b. There are a stable limit cycle and an unstable fixed point in the left-hand side and a stable fixed point on the right-hand side, and there is a saddle-point in the origin. Consequently we observe bistability between self-oscillations and a stable stationary state, and it depends upon the initial conditions whether they are chosen in the basin of attraction of either the limit cycle or the stable fixed point. In the presence of large time delay $\tau$, i.e., if the delay time is much larger than the characteristic fast response time, but on the other hand much smaller than the integral time of the slow variable, the dynamics of the system (\ref{eq1}) becomes completely different. The time delay induces new dynamic regimes, depending upon the system parameters. 

%%%%%%%%%%%%%%%%%%%%%%FIG1%%%%%%%%%%%%%%%%%%%%%%%%%%%%%
\begin{figure}[]
\begin{center}
\includegraphics[width=0.45\textwidth]{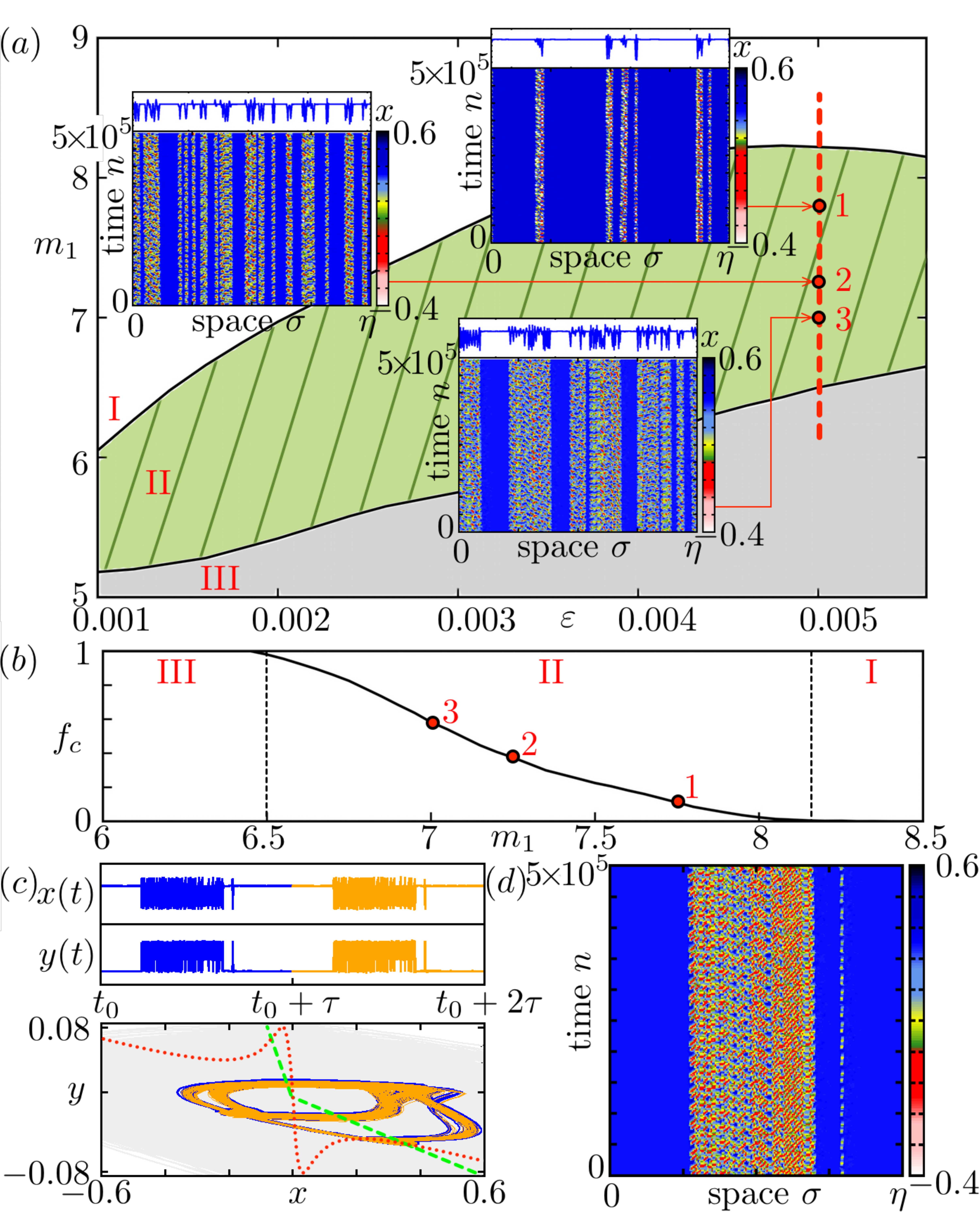}
\end{center}
\caption{(Color online) (a) Map of regimes in the ($\varepsilon, m_{1}$) parameter plane: I (quiescent), II (partially chaotic), III (completely chaotic). Insets: pseudo space-time plots of $x(t)$ at points 1, 2, 3 in regime II and exemplary time series. (b) Dependence of the fraction $f_c$ of the chaotic intervals upon $m_1$. For each value of $m_{1}$ 20 realizations with random initial conditions are averaged. (c) Chimera state for specially prepared initial conditions in point 3. Time series of $x(t)$ and $y(t)$ and phase portrait in the ($x,y$) plane. The $\dot{x}$ and $\dot{y}$ nullclines are shown in red (dotted) and green (dashed), respectively; (d) Space-time plot corresponding to panel (c). Parameters (unless varied in panels (a), (b)): $\varepsilon = 0.005, g = 0.1, a = 200,b = 0.2, m_{1} = 7, m_{2}=1, \tau=200$, and $\eta=200.204045$ (point 1), 200.202711 (point 2), 200.200667 (point 3). Transients\protect\footnotemark[1] of $n=5 \times 10^{6}$ were discarded.}
\label{fig1}
\end{figure}
%%%%%%%%%%%%%%%%%%%%%%%%%%%%%%%%%%%%%%%%%%%%%%%

\section{Coherence-incoherence scenarios}
We will now consider a virtual space-time representation of the delayed-feedback system (\ref{eq1}). In this way the purely temporal dynamics can be mapped onto space-time ($\sigma,n$) \cite{LAR13,LAR15,ARE92,GIA96,MAR12a} 
by introducing $t=n\eta+\sigma$ with an integer (slow) time variable $n$, and a pseudo-space variable $\sigma \in [0,\eta]$, where $\eta=\tau+\delta$ with a small quantity $\delta$, of the order O($\varepsilon / \tau$), which is due to the finite internal response time of the system. For each set of parameters a unique value $\eta$ can be chosen such that the oscillatory dynamics is periodic with period $\eta$. 

The most pronounced changes of the behavior are observed by tuning the parameters $m_1$ (which controls the position of the unstable fixed point on the nonlinear characteristic, see Fig.~\ref{fig4}b) and $\varepsilon$. The resulting dynamic regimes are depicted in Fig.~\ref{fig1}a. \footnotetext[1]{It is important to note that the periodicity $\eta= \tau + \delta$ also changes during the transients and reaches its asymptotic value only after long transients.}They include a completely quiescent steady state (white area I), and a regime of fully developed spatio-temporal chaos (gray area III). With decreasing $m_1$ a transition from the quiescent regime to a spatio-temporal chaos regime occurs by passing through the region II (green hatched area), in which the quiescent behavior alternates with chaotic dynamics. The transition starts with the appearance of {\it solitary states}, which become more and more frequent as $m_1$ decreases from points 1 to 3, see insets of Fig.~\ref{fig1}a. The fraction $f_c$ of the chaotic intervals with respect to the total interval $(0,\eta)$ increases gradually from zero to one, as $m_1$ passes through the partially chaotic regime II (Fig.~\ref{fig1}b). This is a coherence-incoherence scenario distinct from the conventional chimera scenario, which is characterized by gradually growing width of compact intervals of incoherent dynamics embedded in the coherent state. Here, in contrast, we find that the number of small chaotic intervals grows in a non-compact manner, which is similar to the solitary states found in the nonlocally coupled Kuramoto model with inertia \cite{JAR15}, and is familiar from desynchronization transitions, e.g., in Josephson junction arrays \cite{WIE96a}. The solitary states are a manifestation of {\it spatial chaos} in the delayed-feedback system (1), which is characterized by a huge multistability of the states depending on the initial conditions (see \cite{OME11} and references therein). They are reminiscent 
 of space-time patterns of {\em salt-and-pepper instabilities} which occur in spatially extended reaction-diffusion systems with nonlocal spatial coupling \cite{BAC14} in the short wavelength ($k \to \infty$) limit and have been associated with morphogenesis when differentiated cells inhibit the differentiation of neighboring cells, as is seen, for example, with differentiated neuroprogenitor cells in the epithelium of Drosophila embryos \cite{KON10}.

It is also possible to observe chimera states in the transition region II in Fig.~\ref{fig1}a, but only for specially prepared initial conditions (see Appendix). In that case the chaotic segments become localized in compact clusters, and these clusters persist for a long time (up to $t=10^{9}$ time units of our simulations). This regime is characterized by periodic alternation of phases of regular and chaotic dynamics in the $x(t)$ and $y(t)$ time series (Fig.~\ref{fig1}c).  The approximate period of this alternating sequence is close to $\tau$. In the $(x,y)$ coordinates it corresponds to phase trajectories which consist of chaotic and quasi-stationary parts close to some fixed point in phase space. The corresponding space-time plot (Fig.\ref{fig1}d) is analogous to those presented in \cite{LAR13,LAR15}. It consists of two parts: smooth plateaus with almost constant amplitude and oscillatory parts where the dynamics is chaotic. It can be interpreted as two clusters of oscillators continuously distributed in the pseudo-space $\sigma$. Therefore this regime represents a {\em chimera state}. In contrast to the Ikeda model with {\em positive} delayed feedback \cite{LAR13}, the oscillatory dynamics does not include the slow motion along the nullclines (as illustrated in the phase portrait in Fig.~\ref{fig1}c). In the presence of weak additive noise our simulations show that the chimera behavior remains robust. 

In conclusion, the delayed feedback oscillator with negative feedback exhibits two distinct coherence-incoherence scenarios depending upon initial conditions: a novel solitary scenario, and the conventional chimera scenario only for specially prepared initial conditions. 

%%%%%%%%%%%%%%%%%%%%%%FIG2%%%%%%%%%%%%%%%%%%%%%%%%%%%%%
\begin{figure}[]
\begin{center}
\includegraphics[width=0.45\textwidth]{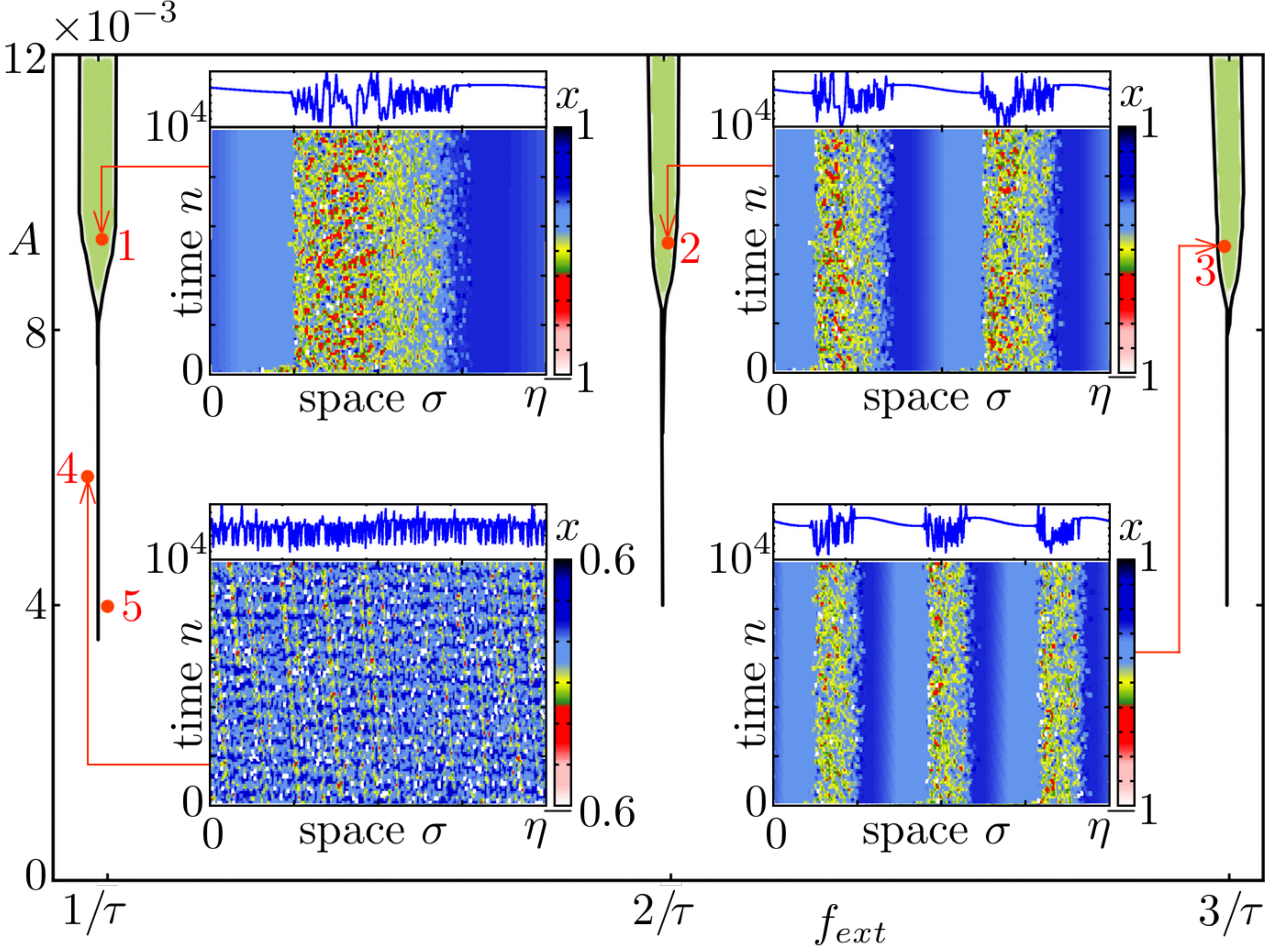}
\end{center}
\caption{(Color online) Deterministic forcing: Map of dynamic regimes in the ($A,f_{ext}$) plane. Insets show pseudo space-time plots and exemplary time series corresponding to points 1 -- 4. Parameters: $\varepsilon = 0.005, g = 0.1, a = 200, b = 0.2, m_{1} = 7, m_{2}=1, \tau=200$, and $\eta =200$ (points 1 -- 3), $\eta =200.19947$ (point 4).}
\label{fig2}
\end{figure}
%%%%%%%%%%%%%%%%%%%%%%%%%%%%%%%%%%%%%%%%%%%%%%%

\section{Deterministic forcing of chimeras} The main focus of this paper is the possibility of controlling the delayed-feedback chimeras by external forcing $F_{ext}(t)$:
\begin{equation}
\label{eq2}
\varepsilon \dot{x} = -y-gx-f(x(t-\tau))-F_{ext}(t), \quad
\dot{y}=x-Q(y)
\end{equation}

First we consider deterministic periodic forcing $F_{ext}(t)=A \sin(2\pi f_{ext} t)$ with amplitude $A$ and frequency $f_{ext}$. A map of the dynamic regimes in the ($A$,$f_{ext}$) plane is shown in Fig. \ref{fig2}. We choose a set of system parameters which corresponds to the solitary state regime II in Fig.\ref{fig1} (corresponding to point 3). The external forcing with small amplitude (points 4,5 in Fig. \ref{fig2}) leads to the suppression of the internal solitary dynamics and gives rise to fully developed chaotic behavior. Surprizingly, an increase of the amplitude $A$ can induce chimera states in a certain region of the parameter plane in which the period of alternation of regular and chaotic dynamics (which is initially equal to $\eta$) is entrained and becomes equal to the period of the external forcing $\eta = \tau= 1/f_{ext}$, i.e., the dynamics of the system (\ref{eq2}) becomes locked to the frequency of the external driving (point 1 in Fig. \ref{fig2}). The regions which correspond to the chimera states resemble Arnold tongues. However, while conventional Arnold tongues arise in systems with two frequencies (either a periodically forced oscillator or two coupled oscillators) when the frequency ratio equals a small rational number, the discussion here is related to the continuous pseudo-space behavior, and we find some novel kind of resonance between the number of incoherent clusters and the external forcing frequency. Outside these tongues, the chimera state is destroyed and we observe spatiotemporal chaos or completely synchronous behavior (in case of large amplitudes $A$).
When the frequency $f_{ext}$ of the driving force is close to $k/\tau$ ($k \in \mathbb{N}$), multi-chimeras with $k=\tau f_{ext}$ incoherent clusters are induced. The locking regimes of these multi-chimeras resemble higher-order Arnold tongues (see the $k=2$ and the $k=3$ tongues around points 2 and 3, respectively, in Fig. \ref{fig2}). Interestingly, this means that multi-chimeras can be induced simply by increasing the driving frequency $f_{ext}$.
%The number of clusters is determined by the frequency ratio $k=\tau f_{ext}$.

%%%%%%%%%%%%%%%%%%%%%%FIG3%%%%%%%%%%%%%%%%%%%%%%%%%%%%%
\begin{figure}[]
\begin{center}
\includegraphics[width=0.45\textwidth]{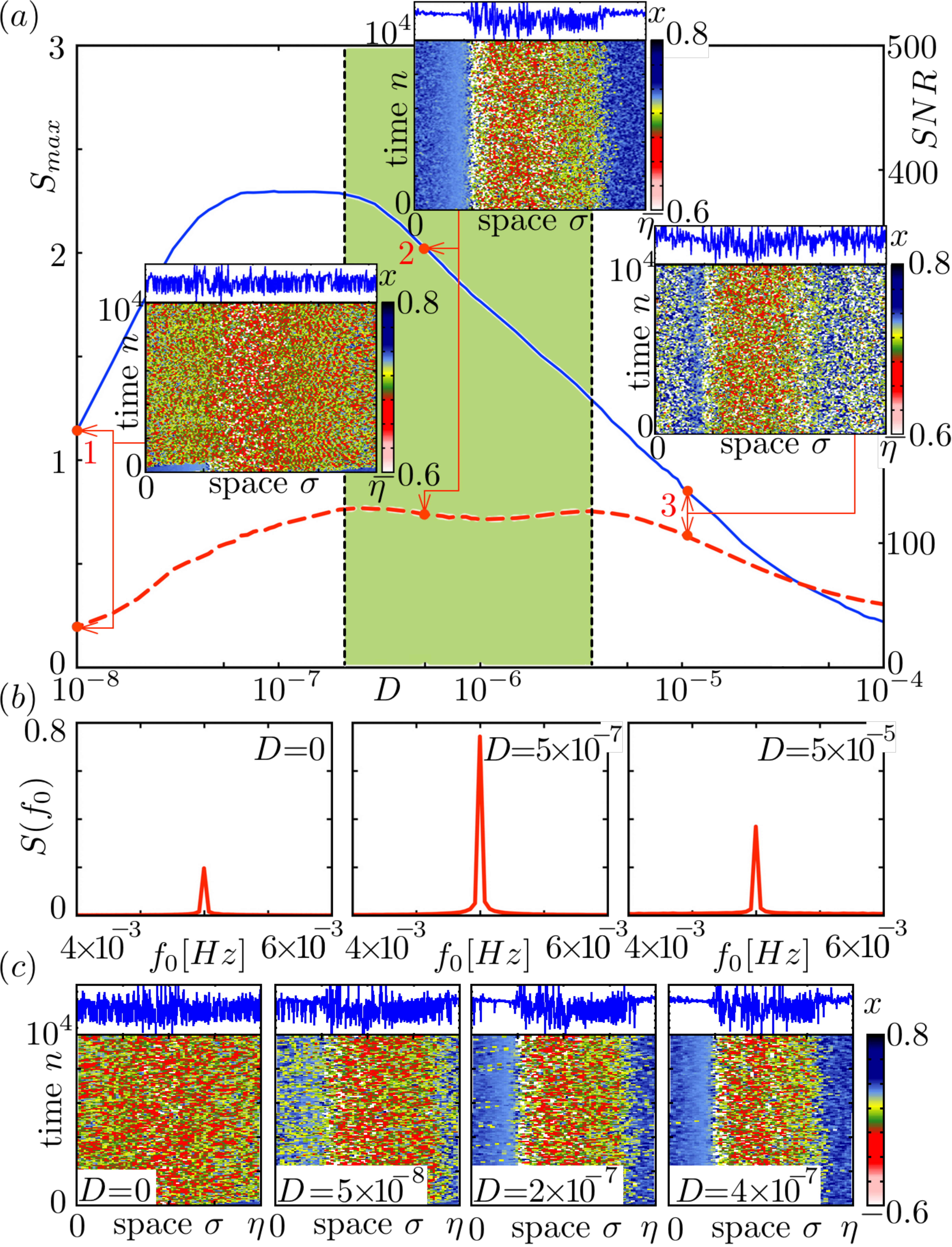}
\end{center}
\caption{(Color online) Delayed-feedback oscillator under stochastic forcing: (a) Peak heights of power spectrum $S_{max}$ of $x(t)$ (red dashed)and signal-to-noise ratio (SNR, blue solid) vs. the noise intensity $D$. Insets: pseudo space-time plots of $x(t)$ and 
exemplary time series at points (1) $D=10^{-8}$; (2) $D=5 \times 10^{-7}$; (3) $D=10^{-5}$. (b) Power spectrum $S(f)$ for increasing noise intensity $D$. (c) Space-time plots of $x(t)$ and exemplary time series for the chimera scenario with increasing $D$.
Parameters: $\varepsilon = 0.005, g = 0.1, a = 200, b = 0.2, m_{1} = 7, m_{2}=1, \tau=\eta=200, A=0.004, f_{ext}=0.005$. }
\label{fig3}
\end{figure}
%%%%%%%%%%%%%%%%%%%%%%%%%%%%%%%%%%%%%%%%%%%%%%%

\section{Stochastic resonance of chimeras} 

Our simulations show that weak additive noise does not essentially affect the observed phenomena, neither in the autonomous system Eq.~(\ref{eq1}) nor in the forced system Eq.~(\ref{eq2}). In both cases, additive noise of large intensity has a destructive character. But if the external forcing $F_{ext}(t)$ of fixed frequency has a stochastic additive component in the forcing amplitude, the presence of noise can play a constructive role similar to what is known from the stochastic resonance of periodically forced systems under the influence of noise.
In this way the favorable action of a periodic forcing term on the generation of chimera behavior, as discussed in the previous section,
can be enhanced by noise if an appropriate resonance condition between the forcing frequency and the noise intensity is met. Such novel
behavior extends the phenomenon of stochastic resonance in a non-delayed system under periodic forcing to chimera behavior of a forced delayed-feedback oscillator; it can be quantified by the power spectral properties, i.e., an enhanced signal-to-noise ratio and increased spectral peak, as in the conventional stochastic resonance.

We add a stochastic term to the periodic forcing $F(t)$ which modulates the external forcing amplitude by Gaussian white noise: 
\begin{equation}
\label{eq3}
F_{ext}(t)=(A+\sqrt{2D}\xi(t))\sin(2\pi f_{ext}t), 
\end{equation}
where $\xi(t)$ is normalized Gaussian white noise $\langle \xi(t) \rangle = 0$, $\langle \xi(t)\xi(t+\tau) \rangle = \delta(\tau)$, and $D$ is the noise intensity. Here we set $f_{ext}=0.005$ and $A=0.004$, which corresponds to spatio-temporal chaos as induced by the external forcing in the deterministic case (see point 5 in Fig.~\ref{fig2}). In case of weak noise the dynamics of the system (\ref{eq2}) remains chaotic, as shown in the space-time plot for point 1 in Fig. \ref{fig3}a. However, increasing the noise intensity $D$ leads to revival of the chimera (see space-time plot for point 2 in Fig. \ref{fig3}a): there is an optimal noise intensity for which the space-time pattern of the system Eq.~(\ref{eq2}) strongly resembles a deterministic chimera state. One can distinguish alternating sequences of regular and of chaotic dynamics in the time series (upper panel of the inset), although both appear slightly noisy. The corresponding pseudo space-time plot also shows the coexisting regular and chaotic domains (lower panel of inset). Further increase of the noise intensity destroys the noise-induced chimera (see space-time plot for point 3 in Fig.~\ref{fig3}a). This phenomenon of constructive influence of noise in a periodically driven nonlinear system is similar to stochastic resonance \cite{GAM98,ANI99}. Indeed, the noise-induced formation of chimera states is accompanied by an increase of the peak of the power spectrum $S_{max}$ at the resonance frequency $f_{ext}$ (Fig.~\ref{fig3}a,b), followed by a decrease upon further increase of the noise intensity. 
The optimum noise window for observation of the chimera state is marked by green shading in Fig.~\ref{fig3}a).
Such non-monotonic behavior as a function of noise intensity is also found in the signal-to-noise ratio (SNR, see Appendix) (Fig.~\ref{fig3}a). The revival of the chimera states occurs after passing through the point of maximum SNR. The scenario leading from completely incoherent chaotic dynamics at $D=0$ to the noise-induced chimera state at $D=4\times 10^{-7}$ is shown in more detail in Fig.~\ref{fig3}c, it clearly shows how noise increasingly suppresses the chaotic dynamics in the coherent domain of the chimera state.

\section{Conclusion} We have studied a delayed-feedback oscillator model with negative feedback under external deterministic and stochastic forcing. The dynamics is strikingly different from the Ikeda model, which is also a delayed-feedback oscillator, but with opposite sign of the delayed feedback. A scenario for the transition from complete coherence to complete incoherence via salt-and-pepper like solitary states has been identified when the nonlinearity parameter of the oscillator is varied. Further, we have shown that chimera states with controllable characteristics, e.g., a desired number of incoherent clusters can be induced by using external periodic forcing. A generalized form of synchronization with the driving signal leads to Arnold tongues of multi-chimera states when the driving frequency obeys a resonance condition, independently of initial conditions. We have also shown that noise can play a constructive role for controlling the chimera state. Noisy modulation of the external forcing amplitude can induce chimeras in regimes where they do not exist without noise; this is reminiscent of stochastic resonance. Since we have used a simple paradigmatic delayed-feedback oscillator model, our results seem to be applicable to a wide range of delay systems, e.g., in optics and electronics, as well as in other fields where nonlinear delayed-feedback can play a role.\\

This work was supported by DFG in the framework of SFB 910 and by the Russian Foundation for Basic Research (RFBR) (grant No. 15-02-02288). We are very grateful to L. Larger, V. Anishchenko, and T. Vadivasova for helpful discussions. Y.M. and V.S. acknowledge support and hospitality of TU Berlin. 

\section{Appendix}

\textsl{Numerical procedure} -- In all numerical simulations the Heun method with time step $\Delta t =0.005$ was used. We have used random initial conditions in the interval of delay. An exception is Fig.~\ref{fig1}c,d, where Eq.~(\ref{eq1}) was initially modeled with external forcing, using parameters corresponding to a one-cluster chimera state (point 3 in Fig.\ref{fig1}). Then Eq.~(\ref{eq1}) was integrated without external forcing using this chimera state as initial condition.

%%%%%%%%%%%%%%%%%%%%%%FIG5%%%%%%%%%%%%%%%%%%%%%%%%%%%%%
\begin{figure}[]
\begin{center}
\includegraphics[width=0.45\textwidth]{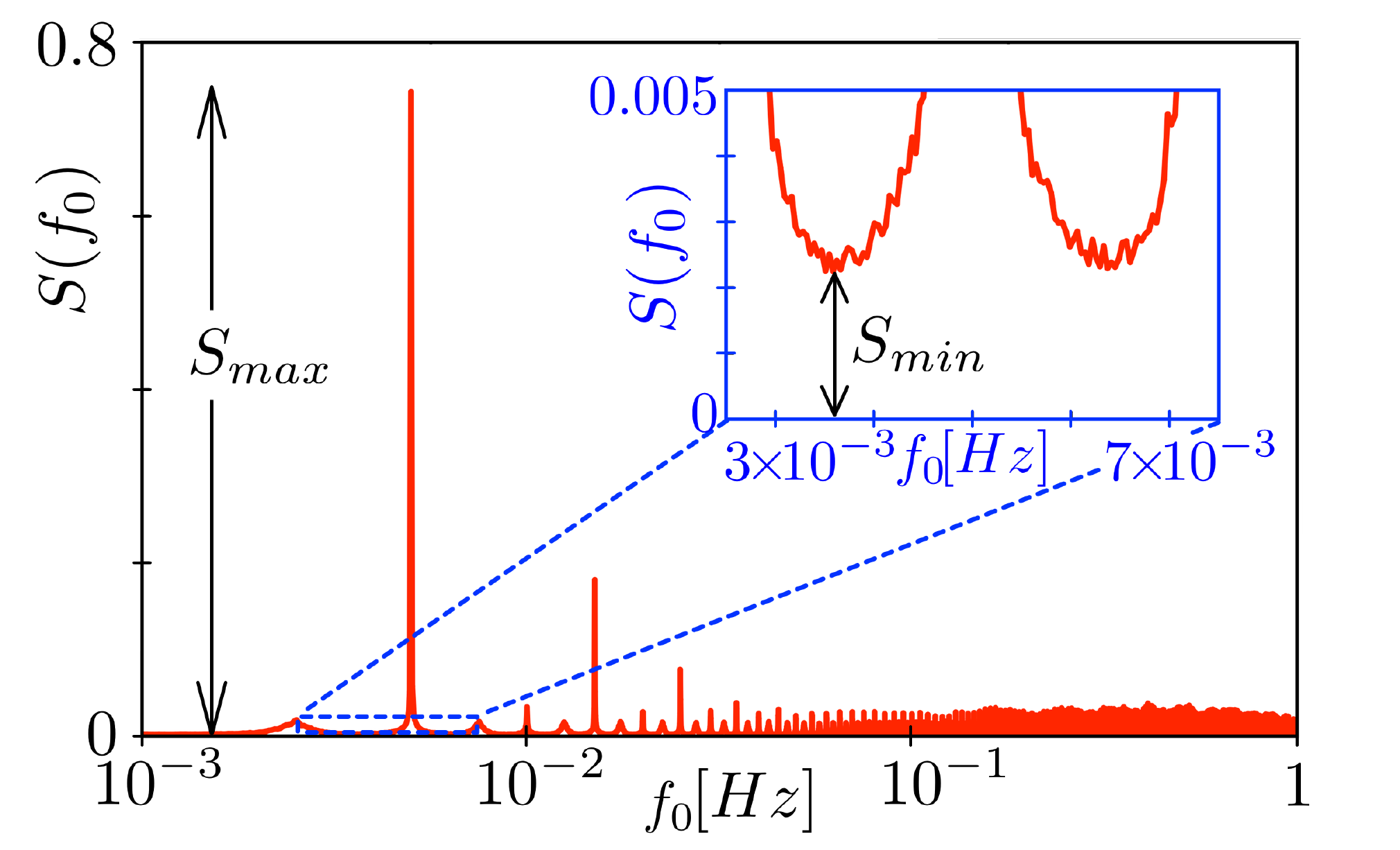}
\end{center}
\caption{Power spectrum $S(f)$ of $x(t)$ of the noise-induced chimera states (Eq.~(\ref{eq2}) with external force Eq.~(\ref{eq3})). The inset shows a blow-up of the parts near the two minima between the main peak, which defines $S_{min}$. Parameters: $\varepsilon = 0.005, g = 0.1, a = 200, b = 0.2, m_{1} = 7, m_{2}=1, \tau=200, A=0.004, f_{ext}=0.005, D=5\times 10^{-7}$.
}
\label{fig5}
\end{figure}
%%%%%%%%%%%%%%%%%%%%%%%%%%%%%%%%%%%%%%%%%%%%%%%

\textsl{Signal-to-noise ratio} -- Here we will describe the method of calculation of the signal-to-noise ratio, which is a common measure for stochastic resonance. The power spectrum for chimera states under stochastic driving for the parameters of Fig.~\ref{fig3} and an optimum noise intensity $D$ is depicted in Fig.~\ref{fig5}. It includes the spectral peak $S_{max}$ at the frequency of external forcing, which is also the main peak in the power spectrum. The power spectrum also has a minimum $S_{min}$ close to the spectral peak. In radiophysics the most common definition of the signal-to-noise ratio (SNR) is $SNR = P_{S}/P_{N}$, where $P_{S}$ is the power of the signal and $P_{N}$ is the noise power. The following formula of SNR corresponds to the harmonic external input signal in experiments: $SNR=H_{s}/H_{n}$, where  $H_{s}$ is the height of the spectral line above the background noise level in the power spectrum, and $H_{n}$ is the background noise level close to the resonance frequency $f_{ext}$, and thus in terms of the power spectrum $SNR=\frac{S_{max}-S_{min}}{S_{min}}$. 

%%\bibliography{sem15b_EPL}
%\bibliography{ref}  

\begin{thebibliography}{10}
\expandafter\ifx\csname url\endcsname\relax\def\url#1{\texttt{#1}}\fi

\bibitem{KUR02a}
\Name{Kuramoto Y. \and Battogtokh D.} \REVIEW{Nonlin. Phen. in Complex
  Sys.}{5}{2002}{380}.

\bibitem{ABR04}
\Name{Abrams D.~M. \and Strogatz S.~H.}
  \REVIEW{Phys.~Rev.~Lett.}{93}{2004}{174102}.

\bibitem{MOT10}
\Name{Motter A.~E.} \REVIEW{Nature Phys.}{6}{2010}{164}.

\bibitem{PAN15}
\Name{Panaggio M.~J. \and Abrams D.~M.} \REVIEW{Nonlinearity}{28}{2015}{R67}.

\bibitem{HAG12}
\Name{Hagerstrom A.~M., Murphy T.~E., Roy R., H{\"o}vel P., Omelchenko I. \and
  Sch{\"o}ll E.} \REVIEW{Nature Phys.}{8}{2012}{658}.

\bibitem{TIN12}
\Name{Tinsley M.~R., Nkomo S. \and Showalter K.} \REVIEW{Nature
  Phys.}{8}{2012}{662}.

\bibitem{MAR13}
\Name{Martens E.~A., Thutupalli S., Fourriere A. \and Hallatschek O.}
  \REVIEW{Proc. Nat. Acad. Sciences}{110}{2013}{10563}.

\bibitem{WIC13}
\Name{Wickramasinghe M. \and Kiss I.~Z.} \REVIEW{PLoS ONE}{8}{2013}{e80586}.

\bibitem{SCH14a}
\Name{Schmidt L., Sch{\"o}nleber K., Krischer K. \and Garcia-Morales V.}
  \REVIEW{Chaos}{24}{2014}{013102}.

\bibitem{ROS14a}
\Name{Rosin D.~P., Rontani D., Haynes N.~D., Sch{\"o}ll E. \and Gauthier D.~J.}
  \REVIEW{Phys. Rev.~E}{90}{2014}{030902(R)}.

\bibitem{KAP14}
\Name{Kapitaniak T., Kuzma P., Wojewoda J., Czolczynski K. \and Maistrenko Y.}
  \REVIEW{Sci. Rep.}{4}{2014}{6379}.

\bibitem{GAM14}
\Name{Gambuzza L.~V., Buscarino A., Chessari S., Fortuna L., Meucci R. \and
  Frasca M.} \REVIEW{Phys. Rev. E}{90}{2014}{032905}.

\bibitem{SIE14c}
\Name{Sieber J., Omel'chenko O. \and Wolfrum M.} \REVIEW{Phys. Rev.
  Lett.}{112}{2014}{054102}.

\bibitem{BIC15}
\Name{Bick C. \and Martens E.~A.} \REVIEW{New J.~Phys.}{17}{2015}{033030}.

\bibitem{OME16}
\Name{Omelchenko I., Omel'chenko O., Zakharova A., Wolfrum M. \and Sch{\"o}ll
  E.} \REVIEW{Phys. Rev. Lett.}{116}{2016}{114101}.

\bibitem{RAT00}
\Name{Rattenborg N.~C., Amlaner C.~J. \and Lima S.~L.} \REVIEW{Neurosci.
  Biobehav. Rev.}{24}{2000}{817}.

\bibitem{LAI01}
\Name{Laing C.~R. \and Chow C.~C.} \REVIEW{Neural Computation}{13}{2001}{1473}.

\bibitem{SAK06a}
\Name{Sakaguchi H.} \REVIEW{Phys. Rev.~E}{73}{2006}{031907}.

\bibitem{ROT14}
\Name{Rothkegel A. \and Lehnertz K.} \REVIEW{New J. of
  Phys.}{16}{2014}{055006}.

\bibitem{MOT13a}
\Name{Motter A.~E., Myers S.~A., Anghel M. \and Nishikawa T.} \REVIEW{Nature
  Phys.}{9}{2013}{191}.

\bibitem{GON14}
\Name{Gonzalez-Avella J.~C., Cosenza M.~G. \and Miguel M.~S.}
  \REVIEW{Physica~A}{399}{2014}{24}.

\bibitem{HU93a}
\Name{Hu G., Ditzinger T., Ning C.~Z. \and Haken H.}
  \REVIEW{Phys.~Rev.~Lett.}{71}{1993}{807}.

\bibitem{PIK97}
\Name{Pikovsky A. \and Kurths J.} \REVIEW{Phys.~Rev.~Lett.}{78}{1997}{775}.

\bibitem{LIN99a}
\Name{Lindner B. \and Schimansky-Geier L.}
  \REVIEW{Phys.~Rev.~E}{60}{1999}{7270}.

\bibitem{LIN04}
\Name{Lindner B., Garc{\'i}a-Ojalvo J., Neiman A.~B. \and Schimansky-Geier L.}
  \REVIEW{Phys.~Rep.}{392}{2004}{321}.

\bibitem{GAM98}
\Name{Gammaitoni L., H{\"a}nggi P., Jung P. \and Marchesoni F.} \REVIEW{Rev.
  Mod. Phys.}{70}{1998}{223}.

\bibitem{ANI99}
\Name{Anishchenko V.~S., Neiman A.~B., Moss F. \and Schimansky-Geier L.}
  \REVIEW{Phys. Usp.}{42}{1999}{7}.

\bibitem{LOO16}
\Name{Loos S., Claussen J.~C., Sch{\"o}ll E. \and Zakharova A.} \REVIEW{Phys.
  Rev. E}{93}{2016}{012209}.

\bibitem{LAR13}
\Name{Larger L., Penkovsky B. \and Maistrenko Y.} \REVIEW{Phys. Rev.
  Lett.}{111}{2013}{054103}.

\bibitem{LAR15}
\Name{Larger L., Penkovsky B. \and Maistrenko Y.} \REVIEW{Nature
  Commun.}{6}{2015}{7752}.

\bibitem{SCH03h}
\Name{Schanz M. \and Pelster A.} \REVIEW{SIAM J. Appl. Dyn.
  Syst.}{2}{2003}{277}.

\bibitem{HIZ07}
\Name{Hizanidis J., Aust R. \and Sch{\"o}ll E.}
  \REVIEW{Int.~J.~Bifur.~Chaos}{18}{2008}{1759}.

\bibitem{PYR92}
\Name{Pyragas K.} \REVIEW{Phys. Lett.~A}{170}{1992}{421}.

\bibitem{HOE05}
\Name{H{\"o}vel P. \and Sch{\"o}ll E.} \REVIEW{Phys.~Rev.~E}{72}{2005}{046203}.

\bibitem{ARE92}
\Name{Arecchi F.~T., Giacomelli G., Lapucci A. \and Meucci R.}
  \REVIEW{Phys.~Rev.~A}{45}{1992}{R4225}.

\bibitem{GIA96}
\Name{Giacomelli G. \and Politi A.} \REVIEW{Phys.~Rev.~Lett.}{76}{1996}{2686}.

\bibitem{IKE79}
\Name{Ikeda K.} \REVIEW{Opt. Commun.}{30}{1979}{257}.

\bibitem{MAR12a}
\Name{Martinenghi R., Rybalko S., Jacquot M., Chembo Y.~K. \and Larger L.}
  \REVIEW{Phys. Rev. Lett.}{108}{2012}{244101}.

\bibitem{GIA12}
\Name{Giacomelli G., Marino F., Zaks M.~A. \and Yanchuk S.} \REVIEW{Europhys.
  Lett.}{99}{2012}{58005}.

\bibitem{VRE94}
\Name{van Vreeswijk C., Abbott L.~F. \and Ermentrout G.~B.} \REVIEW{J. Comp.
  Neurol.}{1}{1994}{313}.

\bibitem{LEH11}
\Name{Lehnert J., Dahms T., H{\"o}vel P. \and Sch{\"o}ll E.} \REVIEW{Europhys.
  Lett.}{96}{2011}{60013}.

\bibitem{KEA12}
%\Name{Keane A., Dahms T., Lehnert J., Suryanarayana S.~A., H{\"o}vel P. \and
%  Sch{\"o}ll E.} \REVIEW{Eur. Phys. J.~B}{85}{2012}{407}.
\Name{Keane A. et al.} \REVIEW{Eur. Phys. J.~B}{85}{2012}{407}.

\bibitem{LAD13}
\Name{Ladenbauer J., Lehnert J., Rankoohi H., Dahms T., Sch{\"o}ll E. \and
  Obermayer K.} \REVIEW{Phys.~Rev.~E}{88}{2013}{042713}.

\bibitem{KAN75}
\Name{Kano G., Takagi H. \and Teramoto I.} \REVIEW{Electronics}{}{1975}{105}.

\bibitem{JAR15}
\Name{Jaros P., Maistrenko Y. \and Kapitaniak T.} \REVIEW{Phys. Rev.
  E}{91}{2015}{022907}.

\bibitem{WIE96a}
\Name{Wiesenfeld K., Colet P. \and Strogatz S.~H.} \REVIEW{Phys. Rev.
  Lett.}{76}{1996}{404}.

\bibitem{OME11}
\Name{Omelchenko I., Maistrenko Y., H{\"o}vel P. \and Sch{\"o}ll E.}
  \REVIEW{Phys. Rev. Lett.}{106}{2011}{234102}.

\bibitem{BAC14}
\Name{Bachmair C.~A. \and Sch{\"o}ll E.} \REVIEW{Eur.
  Phys.~J.~B}{87}{2014}{276}.

\bibitem{KON10}
\Name{Kondo S. \and Miura T.} \REVIEW{Science}{329}{2010}{1616}.

\end{thebibliography}
%\bibliographystyle{eplbib}

\end{document}